\documentstyle[12pt,aaspp4]{article}
\lefthead{Short et al.}
\righthead{Massive NLTE Novae Models}

\begin{document}

\title{Massive Multi-species, Multi-level NLTE\\
    Model Atmospheres for\\
    Novae in Outburst}

\author{C. Ian Short and Peter H. Hauschildt}
\affil{Dept. of Physics and Astronomy and Center for 
Simulational Physics, University of Georgia,
    Athens, GA 30602-2451}

\and

\author{E. Baron}
\affil{Dept. of Physics and Astronomy, University of Oklahoma, 
    440 W. Brooks, Rm 131, Norman, OK 73091-0225}

% Notice that each of these authors has alternate affiliations, which
% are identified by the \altaffilmark after each name.  The actual alternate
% affiliation information is typeset in footnotes at the bottom of the
% first page, and the text itself is specified in \altaffiltext commands.
% There is a separate \altaffiltext for each alternate affiliation
% indicated above.

%\altaffiltext{2}{Society of Fellows, Harvard University.} 

% The abstract environment prints out the receipt and acceptance dates
% if they are relevant for the journal style.  For the aasms style, they
% will print out as horizontal rules for the editorial staff to type
% on, so long as the author does not include \received and \accepted
% commands.  This should not be done, since \received and \accepted dates
% are not known to the author.

\begin{abstract}

    We have used our {\sc PHOENIX} multi-purpose model atmosphere code to 
calculate atmospheric models that represent novae in the optically thick
wind phases of their outburst.  We have improved the
treatment of NLTE effects by expanding the number of elements
that are included in the calculations from
15 to 19, and the number of ionization stages from 36 to 87.
The code can now treat a total of 10713 levels and 102646 lines in
NLTE.  
Aluminum, P, K, and Ni are included for the first 
time in the NLTE treatment and most elements now have at least the lowest six
ionization stages included in the NLTE calculation.  We have investigated the
effects of expanded NLTE treatment on the chemical concentration of
astrophysically significant species in the atmosphere, the equilibrium structure
of the atmosphere, and the emergent flux distribution.  Although we have found general 
qualitative agreement with previous, more limited NLTE models,
the expanded NLTE treatment leads to significantly different values 
for the size of many of the NLTE deviations.  In particular, for the hottest
model presented here ($T_{\rm eff}=35\, 000$ K), for which NLTE effects are largest, we find 
that the expanded NLTE treatment
{\it reduces} the NLTE effects for these important variables: \ion{H}{1} concentration,
pressure structure, and emergent far $UV$ flux.  Moreover, we find that
the addition of new NLTE species may greatly affect the concentration of
species that were already treated in NLTE, so that, generally, {\it all}
species that contribute significantly to the $e^-$ reservoir or to the
total opacity, or whose line spectrum overlaps or interlocks with
that of a species of interest, must be treated in NLTE to insure an 
accurate result for any particular species.

\end{abstract}

% The different journals have different requirements for keywords.  The
% keywords.apj file, found on aas.org in the pubs/aastex-misc directory, 
% contains a list of keywords used with the ApJ and Letters.  These are 
% usually assigned by the editor, but authors may include them in their 
% manuscripts if they wish. 


% That's it for the front matter.  On to the main body of the paper.
% We'll only put in tutorial remarks at the beginning of each section
% so you can see entire sections together.

% In the first two sections, you should notice the use of the LaTeX \cite
% command to identify citations.  The citations are tied to the
% reference list via symbolic KEYs.  We have chosen the first three
% characters of the first author's name plus the last two numeral of the
% year of publication.  The corresponding reference has a \bibitem
% command in the reference list below.
%
% Please see the AASTeX manual for a more complete discussion on how to make
% \cite-\bibitem work for you.   

\section{Introduction}\label{sintro}

   Novae in the optically thick wind phase of their outburst pose a special 
challenge for atmospheric modeling
because of their steep temperature gradients, low densities, 
large geometric extents, and high velocity differential flows.
For example, a model with a bolometrically defined effective temperature, $T_{\rm eff}$, of
15\, 000 K will have values of the kinetic temperature, $T_{\rm kin}$, that range from 
4500 to 140\, 000 K,
of the mass density, $\rho$, that range from $3\times 10^{-15}$ to 
$6\times 10^{-9}$ g cm$^{-3}$, 
a geometric extent, $R_{\rm out}/R_{\rm in}$, equal to 120 (this paper),
and a typical maximum expansion velocity, $V_0$, of $\approx 2000$ km s$^{-1}$.  
Because of the low values of $\rho$,
radiative rates exceed collisional rates for many important transitions 
throughout the atmosphere.  The steep $T_{\rm kin}$ gradient and low densities 
allow local regions to be exposed through scattering processes to radiation from distant regions
in the atmosphere where the radiation temperature, $T_{\rm rad}$, differs
greatly from the local value of $T_{\rm kin}$.  Moreover,
massive ultraviolet ($UV$) and optical line blanketing cause the radiation
field to deviate greatly from a Planck ($B_\nu$) distribution.  As a result
of these considerations, non-local thermodynamic equilibrium (NLTE) effects
are severe in these atmospheres and must be accounted for to construct 
accurate models.  The low values of $\rho$ cause the
atmosphere to be translucent ($\tau<1$) at large physical depths, which,
combined with the large absolute value of $\bigtriangledown_{\tau}T$, 
conspires to allow the line spectrum of many ionization stages of an atomic
element to be present in the emergent flux spectrum.  Moreover, the rapid
expansion velocities and steep $v(r)$ gradients cause this rich line spectrum
to be smeared, which complicates the radiative transfer by increasing the
amount of overlap among different transitions.  Also, the large velocities
cause special relativistic effects to be marginally significant in the transfer of radiation, 
at least to first order in $v/c$.  Finally, the large geometric  
extent allows sphericity effects to be important. 

\paragraph{}

   \cite[]{nlte1} and \cite[]{nlte2} (henceforth HSSBSA) 
investigated the effects of NLTE on 
atmospheric models of novae in the optically thick wind phase of their outburst.
They concluded that an accurate treatment of NLTE effects for many transitions
is critical for correctly calculating both the structure of the atmosphere
and the emergent flux spectrum.  Among their main results are that: 
1) NLTE effects change the predicted concentration
of minority ionization stages of elements that contribute significantly to the line 
opacity, such as Fe and CNO, 
by as much as three orders of magnitude at depths where 
the continuum optical depth at $5000$\AA, $\tau_{\rm cont, 5000}$, is less 
than $10^{-2}$,
2) NLTE effects brighten the predicted $UV$ pseudo-continuum 
by as much as an order of magnitude in the case of their $25\, 000$ K
model in the range $\lambda< 1000$\AA,
and 3) a minimal incorporation of NLTE physics by including a large
single scattering
albedo in the line source functions of an LTE synthetic spectrum gives rise
to a spectrum with approximately the same overall level of line blanketing
absorption as a more complete NLTE treatment, although any individual line
profile may be inaccurate. 
HSSBSA concluded
that their NLTE models provided a better fit to observed spectra of novae.

\paragraph{}

   One of the limitations of the modeling of \cite{nlte1}
and HSSBSA is the limited number of elements and
ionization stages treated in NLTE (15 elements and 36 stages).  Given the importance
of NLTE effects, we have increased the NLTE treatment to include 19 elements and 87
ionization stages.  In Section \ref{snlte} we describe the NLTE treatment, in Section 
\ref{smods}
we describe the nova models, in Section \ref{sresult} we present the results of our 
calculations, and in Section \ref{sconc} we present our conclusions. 

\section{NLTE Treatment}\label{snlte}

\begin{deluxetable}{lrrrrrrr}
\small
\tablecaption{Species and number of levels and primary transitions treated
in NLTE.}
\tablecomments{New elements and stages are indicated with bold face text.  Species
for which the model atom or ion has been improved are denoted with italics.}
\tablenotetext{a}{Only 30 levels used in model (see text).}
\label{t1}
\tablecolumns{8}
\tablewidth{0pt}
\tablehead{
\colhead{Element} & \multicolumn{7}{c}{Ionization Stage} \\
\colhead{} & \colhead{\ion{}{1}} & \colhead{\ion{}{2}} & \colhead{\ion{}{3}} & \colhead{\ion{}{4}} & \colhead{\ion{}{5}} & \colhead{\ion{}{6}} & \colhead{\ion{}{7}}   } 
\startdata
{\it H}       & {\it 80/3160}\tablenotemark{a} &\nodata &\nodata &\nodata &\nodata &\nodata &\nodata \\
He      &  19/37 & 10/45 &\nodata &\nodata &\nodata &\nodata &\nodata \\
Li      & 57/333 & {\bf 55/124} &\nodata &\nodata &\nodata &\nodata &\nodata \\
C       &  228/1387 & 85/336 & 79/365 & 35/171  &\nodata &\nodata &\nodata \\
N       & 252/2313 & 152/1110 & 87/266 &80/388 &39/206 & 15/23  &\nodata\\
O       &  36/66 & 171/1304 & 137/765 & 134/415 & 97/452 & 39/196  &\nodata\\
Ne      &  26/37  &\nodata &\nodata &\nodata &\nodata &\nodata &\nodata\\
{\it Na}     &  {\it 53/142} & {\bf 35/171} & {\bf 69/353} & {\bf 46/110} & {\bf 64/187} & {\bf 102/375}  &\nodata\\
{\it Mg}      & {\bf 273/835} & {\it 72/340} & {\bf 91/656} & {\bf 54/169} & {\bf 53/133} & {\bf 78/180}  &\nodata\\
{\bf Al}      & {\bf 111/250} & {\bf 188/1674}&{\bf 58/297} & {\bf 31/142} & {\bf 49/77} & {\bf 40/93}  &\nodata\\
Si      & {\bf 329/1871}& 93/436 & 155/1027&{\bf 52/292} & {\bf 35/125} & {\bf 36/49}  &\nodata \\
{\bf P}       & {\bf 229/903} & {\bf 89/760} & {\bf 51/145} & {\bf 50/174} & {\bf 40/204} & {\bf 10/9}  &\nodata\\
S       & {\bf 146/439} & 84/444 & 41/170 & {\bf 28/50} & {\bf 19/41} & {\bf 31/144}  &\nodata\\
{\bf K}       & {\bf 73/210} & {\bf 22/66} & {\bf 38/178} & {\bf 24/57} & {\bf 29/75}  &\nodata &\nodata\\
Ca      & {\bf 194/1029}& 87/455 & {\bf 150/1661} & {\bf 67/122} & {\bf 39/91} & {\bf 23/37} & {\bf 26/59} \\
Ti      & 395/5279 & 204/2399  &\nodata &\nodata &\nodata &\nodata &\nodata\\
Fe      & 494/6903 & 617/13675&566/9721 & {\bf 243/2592} & {\bf 132/961} & {\bf 87/551}  &\nodata\\
Co      & 316/4428 & 255/2725 & 213/2248  &\nodata &\nodata &\nodata &\nodata\\
{\bf Ni}      & {\bf 153/1690} & {\bf 429/7445} & {\bf 259/3672} & {\bf 189/1845} & {\bf 245/2638} & {\bf 246/2868} &\nodata\\
\\
Total   & 10713/102646
\enddata
\end{deluxetable}

   Table 1 shows the {\em complete} set of chemical elements and
ionization stages now treated in NLTE by {\sc PHOENIX} along with the 
number of levels and primary bound-bound transitions included in the model
atoms and ions.  Primary transitions are those that connect states with
observed energy levels and that have a $\log gf$ value greater than $-3.0$.
They are explicitly treated in the NLTE rate equations.
All other transitions of a species treated in 
NLTE are considered secondary.  Secondary transitions are not included in the 
rate equations, but are still included in the line opacity, with
the departure co-efficients of any levels not included in the rate equations
set equal to that of the ground state (see \cite{hausfe2}
for a detailed description).    
New species are denoted in bold face type and species for
which the treatment has been improved with enlarged model atoms are denoted
with italics.  We have added four new elements and 51 new ionization stages 
to the NLTE treatment with the result that the code 
can now treat a total of 87 ionization stages among 19 elements.  
This has increased the number of levels and lines included
in the statistical equilibrium solution by 5415 and 42521, respectively,
thereby approximately doubling the total numbers included.  Aluminum, P,
K, and Ni are now included in the NLTE treatment for the first time.  The
latter is important for supernova modeling because of the role of the
radio-active decay of Ni in the energy balance of SNe envelopes, but is
inconsequential for the nova modeling described here.  Most elements now have 
at least the lowest six ionization stages treated in NLTE.  For most elements
this level of ionization corresponds to an energy in the range of 100 to 200
eV, which is well above the local thermal energy in the atmospheres of the hottest 
novae that we have modeled.  In addition, we have improved the NLTE treatment of 
\ion{Na}{1} and \ion{Mg}{2} by enlarging the model atoms from three 
levels and two lines to 53 levels and
142 lines for \ion{Na}{1}, and from 18 levels and 37 lines to 273 levels and 835 lines
for \ion{Mg}{2}.  

\paragraph{}

We have also enlarged the \ion{H}{1} model atom from 
30 levels and 435 lines to 80 levels and 3160 lines.  However, we restrict the models
to 30 levels of \ion{H}{1} because our treatment of dissolution effects among
high lying levels is only approximate.  These dissolution 
effects have been shown by \cite{hubeny}
to have an important effect on model structure and synthetic spectra for models of
main sequence stars in the temperature range being investigated here.   We note 
that because Novae atmospheres are expanding rapidly, the gas pressure in the line
forming region is orders of magnitude less than that of a static White Dwarf (WD)
star.  Therefore, we expect dissolution effects to be negligible for states with
principal quantum number, $n$, less than 30.  

\paragraph{}

   Atomic data for the energy levels and $b-b$ transitions have been taken from 
Kurucz (\cite{krczcd22}, \cite{krczcd23}).  An accurate treatment of photo-ionization
is important for the correct solution of the opacity and chemical equilibrium of
a NLTE gas.  We have used the resonance averaged Opacity Project (\cite{op}) 
data of \cite{bfop} for 
the ground state photo-ionization cross 
sections for \ion{Li}{1}-\ion{}{2}, \ion{C}{1}-\ion{}{4}, \ion{N}{1}-\ion{}{6},
\ion{O}{1}-\ion{}{6}, \ion{Ne}{1}, \ion{Na}{1}-\ion{}{6}, \ion{}{1}-\ion{}{6}, 
\ion{Al}{1}-\ion{}{6}, \ion{Si}{1}-\ion{}{6}, \ion{S}{1}-\ion{}{6}, 
\ion{Ca}{1}-\ion{}{7}, and \ion{Fe}{1}-\ion{}{6}.  These data also incorporate
$X$-ray band opacity due to ionizations from the $K$ electron shell.  For those
species that were already included in the {\sc PHOENIX} NLTE treatment, the
use of this $b-f$ opacity data is an improvement.  For the ground states of
all stages of P, Ti, Co, and Ni, and for the {\it excited} states of {\it all} species,
we have used the cross sectional data previously incorporated into {\sc PHOENIX},
which are those of Reilman \& Manson (\cite{oldbf2}) or those compiled by Mathisen 
(\cite{oldbf}).  We account for coupling among all bound 
levels by electronic collisions using cross-sections calculated with 
the formula of \cite{allen}, except
for those levels connected by radiative transitions, for which we use the
formula of \cite{vrm}.  The cross sections of ionizing collisions with electrons
are calculated with the formula of \cite{draw}.  

\paragraph{}

   With the expanded NLTE treatment, {\sc PHOENIX} now treats 
simultaneously $\sim 10000$ levels and $\sim 100\, 000$ primary transitions
in NLTE.  This massive NLTE treatment is made possible by the 
efficiency of the operator splitting
method for solving the multi-level NLTE rate equations (\cite{phxrate}) 
and of the ALI/OS method for solving the radiative transfer 
equation (\cite{phxali}), and by the parallel implementation of 
{\sc PHOENIX} (\cite{prl1}, \cite{prl2}).

\section{Models}\label{smods}

\begin{table}
\caption{Parameters of the nova atmospheres}
\label{t2}
\begin{tabular}{ll|r|r|r|r|r}
Model & Time & $T_{\rm eff}$ (K) & $r(\tau_{\rm cont, 5000}=1)$ (cm) & $L_{\rm bol}$ ($L_\odot$) & $R_{\rm out}/R_{\rm in}$ & $v_0$ (km s$^{-1}$) \\
$M1$  & $t_1$ & $15\, 000$       & $2.3\times 10^{12}$               & $50\, 000$                   & $120$                  & 2000 \\
$M2$  & $t_2$ & $25\, 000$       & $8.3\times 10^{11}$               & $50\, 000$                   & $210$                  & 2000 \\
$M3$  & $t_3$ & $35\, 000$       & $4.2\times 10^{11}$               & $50\, 000$                   & $270$                  & 2000 \\
\end{tabular}
\end{table} 

   We have calculated three models, designated $M1$, $M2$, and $M3$, that represent 
the state of a nova at three different times
during the optically thick wind phase of the outburst.  Table 2 presents
some of the model parameters.  During the
optically thick wind phase, the radius at which $\tau_{\rm cont, 5000}$ equals unity
is decreasing due to the decrease in $\rho$ as the atmosphere expands.  The decreasing 
value of $\rho$ also allows energetic photons
from the central engine to penetrate further out into the expanding atmosphere,
thereby increasing $T_{\rm eff}$.
For these spherically extended atmospheres the value of $T_{\rm eff}$ is defined as the 
model temperature that corresponds to the total frequency integrated luminosity 
($\sigma T_{\rm eff}^4={3\over {4\pi R_{\tau=1}^2}}\int_0^\infty L_\nu d\nu$).  During this phase, 
the photospheric
radius, $r(\tau_{\rm cont, 5000}=1)$, and $T_{\rm eff}$ change such that the
bolometric luminosity, $L_{\rm bol}$, remains constant.  For our models, $L_{\rm bol}$
is equal to $50\, 000 L_\odot$.
The optical depth grid contains 50 points and spans the range from a 
$\log\tau_{\rm cont, 5000}$ value of -10 to three.  Because of the decline in
$\rho$, a fixed grid in $\tau$ space 
corresponds to a contracting grid in physical distance space.
However, note that the inner radius at the bottom of the model, $R_{\rm in}$, decreases 
faster than the outer radius at the top of the model, $R_{\rm out}$, so that the
geometric extent $R_{\rm out}/R_{\rm in}$, and, therefore, the sphericity, of the 
model is increasing with time.  The density law, 
$\rho(r)=\rho_0(r/R_0)^{-n}$, is prescribed and has a value of $n=3$.
The velocity law, $v(r)=v_0(r/R_0)$, corresponds to a constant mass loss rate
($\dot{M}(r)=$constant), and has a value of $v_0$ equal to $2000$ km s$^{-1}$. 

\paragraph{}

\begin{deluxetable}{lrrrrrrr}
\footnotesize
\tablecaption{Species treated in NLTE for each model in the HSSBSA and current calculations}
\tablecomments{HS denotes the modeling of HSSBSA.  $M1$, $M2$, and $M3$ are defined in Table 2. } 
\label{t3}
\tablecolumns{8}
\tablewidth{0pt}
\tablehead{
\colhead{Element} & \multicolumn{7}{c}{Ionization Stage} \\
\colhead{} & \colhead{\ion{}{1}} & \colhead{\ion{}{2}} & \colhead{\ion{}{3}} & \colhead{\ion{}{4}} & \colhead{\ion{}{5}} & \colhead{\ion{}{6}} & \colhead{\ion{}{7}}   } 
\startdata
{\it H} & HS/$M1$/$M2$/$M3$ &\nodata &\nodata &\nodata &\nodata &\nodata &\nodata \\
He      & HS/$M1$/$M2$/$M3$ & HS/$M1$/$M2$/$M3$ &\nodata &\nodata &\nodata &\nodata &\nodata \\
Li      & None & None &\nodata &\nodata &\nodata &\nodata &\nodata \\
C       & HS/$M1$/$M2$/$M3$  & HS/$M1$/$M2$/$M3$ & HS/$M1$/$M2$/$M3$ & HS/$M1$/$M2$/$M3$  &\nodata &\nodata &\nodata \\
N       & $M1$/$M2$/$M3$ & $M1$/$M2$/$M3$ & $M1$/$M2$/$M3$ &$M1$/$M2$/$M3$ & $M3$ & None  &\nodata\\
O       & $M1$/$M2$/$M3$ & $M1$/$M2$/$M3$ & $M1$/$M2$/$M3$ & $M1$/$M2$/$M3$ & $M3$ & None &\nodata\\
Ne      &  $M1$/$M2$/$M3$ &\nodata &\nodata &\nodata &\nodata &\nodata &\nodata\\
{\it Na}     & $M1$/$M2$/$M3$ & $M1$/$M2$/$M3$ & $M1$/$M2$/$M3$ & $M1$/$M2$/$M3$ & None & None  &\nodata\\
{\it Mg}      & $M1$/$M2$ & HS/$M1$/$M2$/$M3$ & $M1$/$M2$/$M3$ & $M1$/$M2$/$M3$ & None & None  &\nodata\\
{\bf Al}      & $M1$/$M2$/$M3$ & $M1$/$M2$/$M3$ & $M1$/$M2$/$M3$ & $M1$/$M2$/$M3$ & None & None  &\nodata\\
Si      & $M1$/$M2$ & $M1$/$M2$/$M3$ & $M1$/$M2$/$M3$ & $M1$/$M2$/$M3$ & $M3$ & None  &\nodata \\
{\bf P}       & $M1$/$M2$ & $M1$/$M2$/$M3$ & $M1$/$M2$/$M3$ & $M1$/$M2$/$M3$ & $M3$ & None  &\nodata\\
S       & $M1$/$M2$ & $M1$/$M2$/$M3$ & $M1$/$M2$/$M3$ & $M1$/$M2$/$M3$ & $M3$ & None  &\nodata\\
{\bf K}       & None & None & None & None & None  &\nodata &\nodata\\
Ca      & $M1$/$M2$& HS/$M1$/$M2$/$M3$ & $M1$/$M2$/$M3$ & $M1$/$M2$/$M3$ & $M3$ & None & None \\
Ti      & None & None  &\nodata &\nodata &\nodata &\nodata &\nodata\\
Fe      & $M1$/$M2$ & HS/$M1$/$M2$/$M3$ & $M1$/$M2$/$M3$ & $M1$/$M2$/$M3$ & $M3$ & None  &\nodata\\
Co      & None & None & None  &\nodata &\nodata &\nodata &\nodata\\
{\bf Ni}      & None & None & None & None & None & None &\nodata\\
\enddata
\end{deluxetable}

   For each of these models we have calculated a converged solution for
the atmospheric structure that is subject to the constraint of radiative
equilibrium for three cases: 1) quasi-LTE treatment for all 
species in which the Boltzmann and Saha distributions are used for the
level populations, and all lines are assumed to have an albedo for single 
scattering of 0.95, 2) NLTE treatment 
for only the HSSBSA species, which are denoted by 
normal typeface in Table 1, and 3)
NLTE treatment of all relevant species presented in
Table 1.  The line albedo used for case 1 was found by
HSSBSA to be necessary to produce synthetic LTE spectra for nova models in 
which the line strengths were approximately realistic.  For case 3,
relevant species are those whose levels have a numerically significant
population anywhere in the atmosphere and whose line spectrum is already 
known to be an important opacity source for determining the atmospheric
structure or the appearance of the emergent spectrum.  This limit was 
placed on the number of species treated in NLTE for the sake of computational
expediency.  Table 3 lists those
species treated in NLTE in each case for each model.  {\it Note that for no
model are {\bf all} the species in Table 1 treated in NLTE.}  In this regard
we note that some of the species in Table 1 are not important for Nova modeling, 
and the facility to treat them in NLTE was added to {\sc PHOENIX} for the
sake of other applications.  For cases 2 and 3, 
the NLTE problem is converged self-consistently with the atmospheric structure.

\section{Results} \label{sresult}

\subsection{NLTE populations} \label{srespop}

\begin{figure}
\plotfiddle{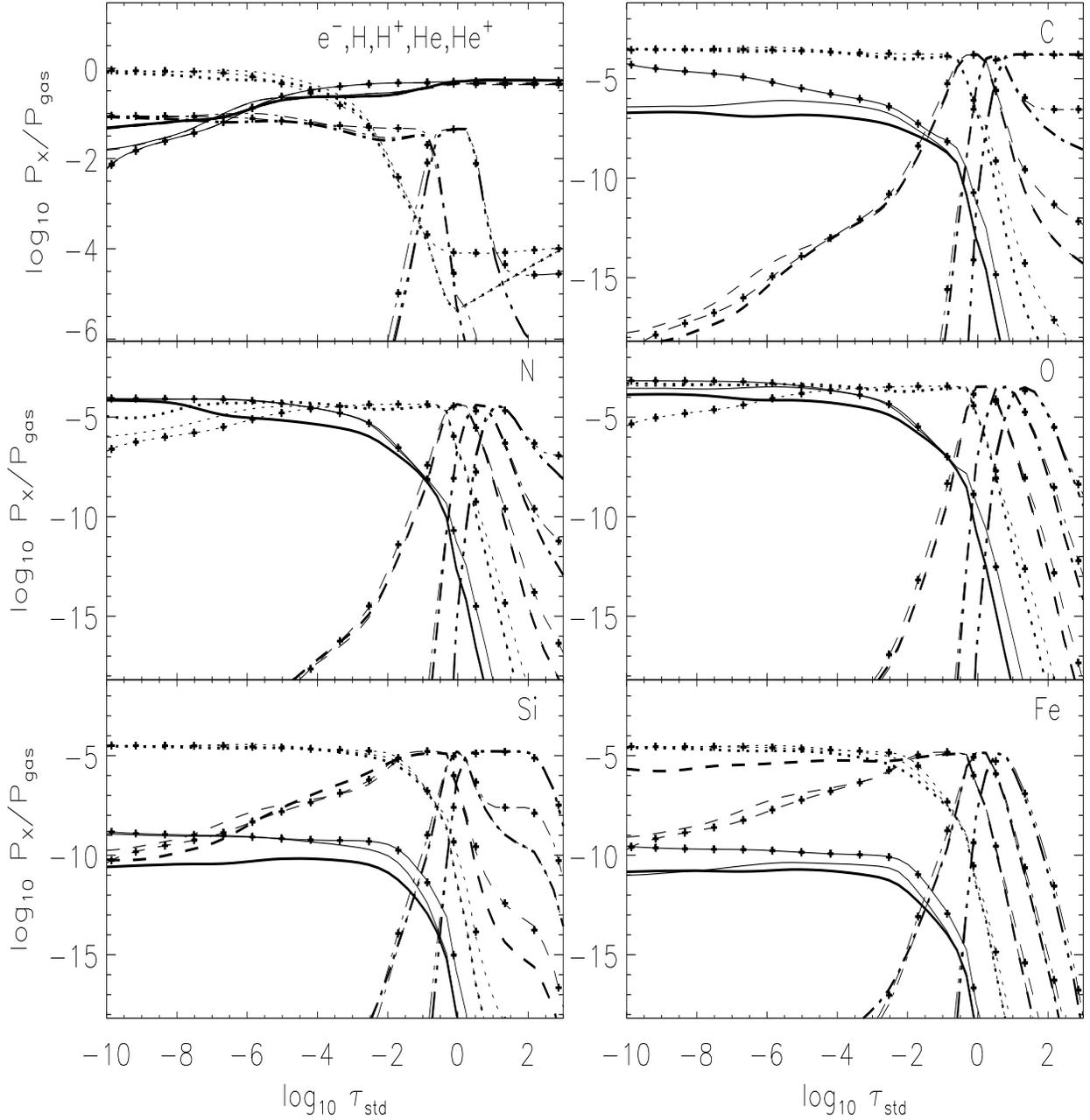}{7.0truein}{0}{100}{135}{-216}{0}
\figcaption[f1.eps]
{Comparison of partial pressure of various species for the \protect $M1$ model.  Thin
lines with \protect $\times$ symbols: LTE, thin lines: HSSBSA NLTE, thick
lines: current NLTE.  Line style
correspond to ionization stages; solid: \protect \ion{}{1}, dot: \protect \ion{}{2}, dash:
\protect \ion{}{3}, dot-dash: \protect \ion{}{4}, dot-dot-dot-dash: \protect \ion{}{5}, except for
upper left panel; solid: \protect $e^-$, dot: \protect \ion{H}{1}, dash: \protect \ion{H}{2},
dot-dash: \protect \ion{He}{1}, dot-dot-dot-dash: \protect \ion{He}{2}. \label{fpp15} } 
\end{figure}

\begin{figure}
\plotfiddle{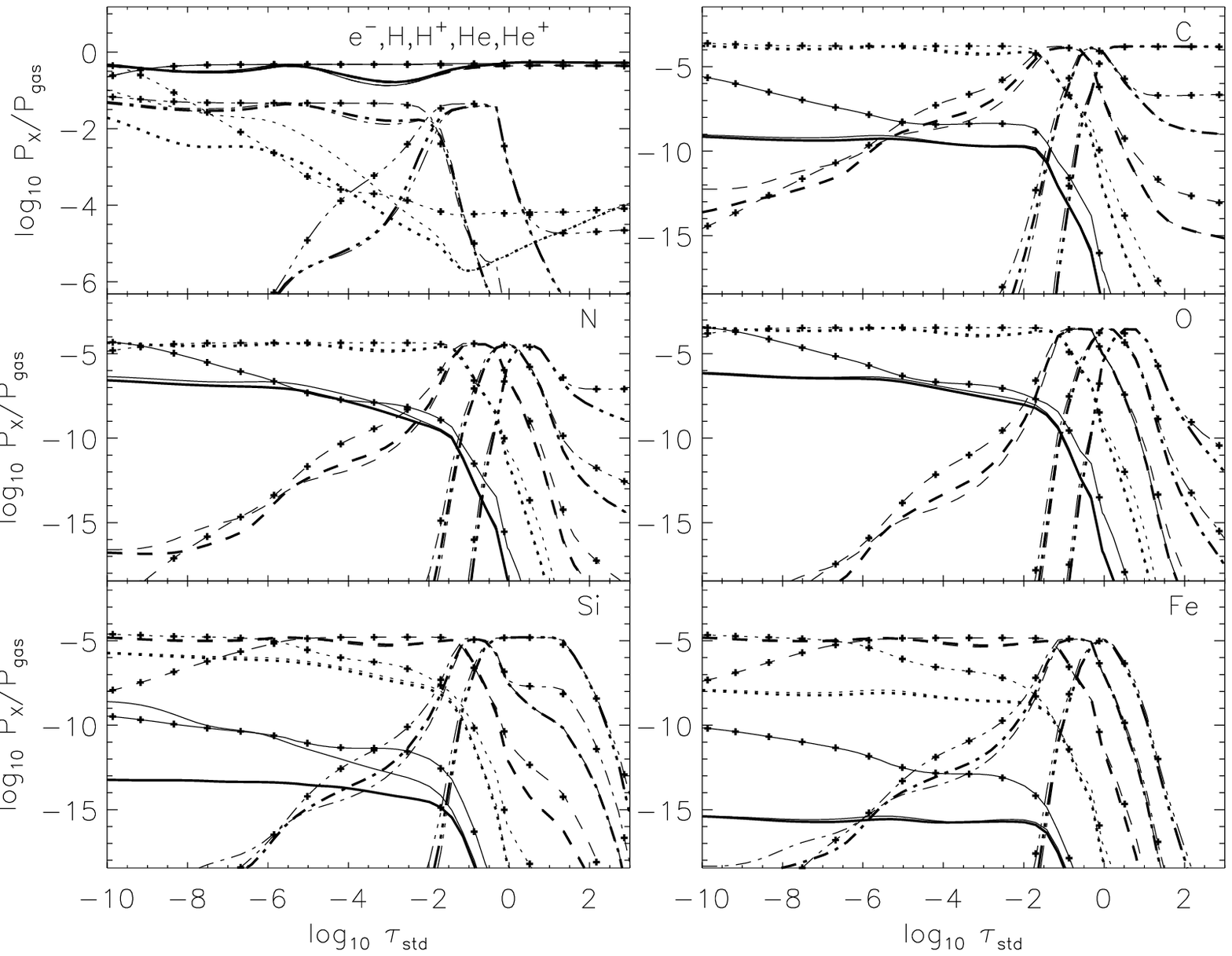}{7.0truein}{0}{100}{135}{-216}{0}
\figcaption[f2.eps]
{Partial pressure of various species for the \protect $M2$ model.  See Fig. \ref{fpp15}. 
\label{fpp25a} }
\end{figure}

\begin{figure}
\plotfiddle{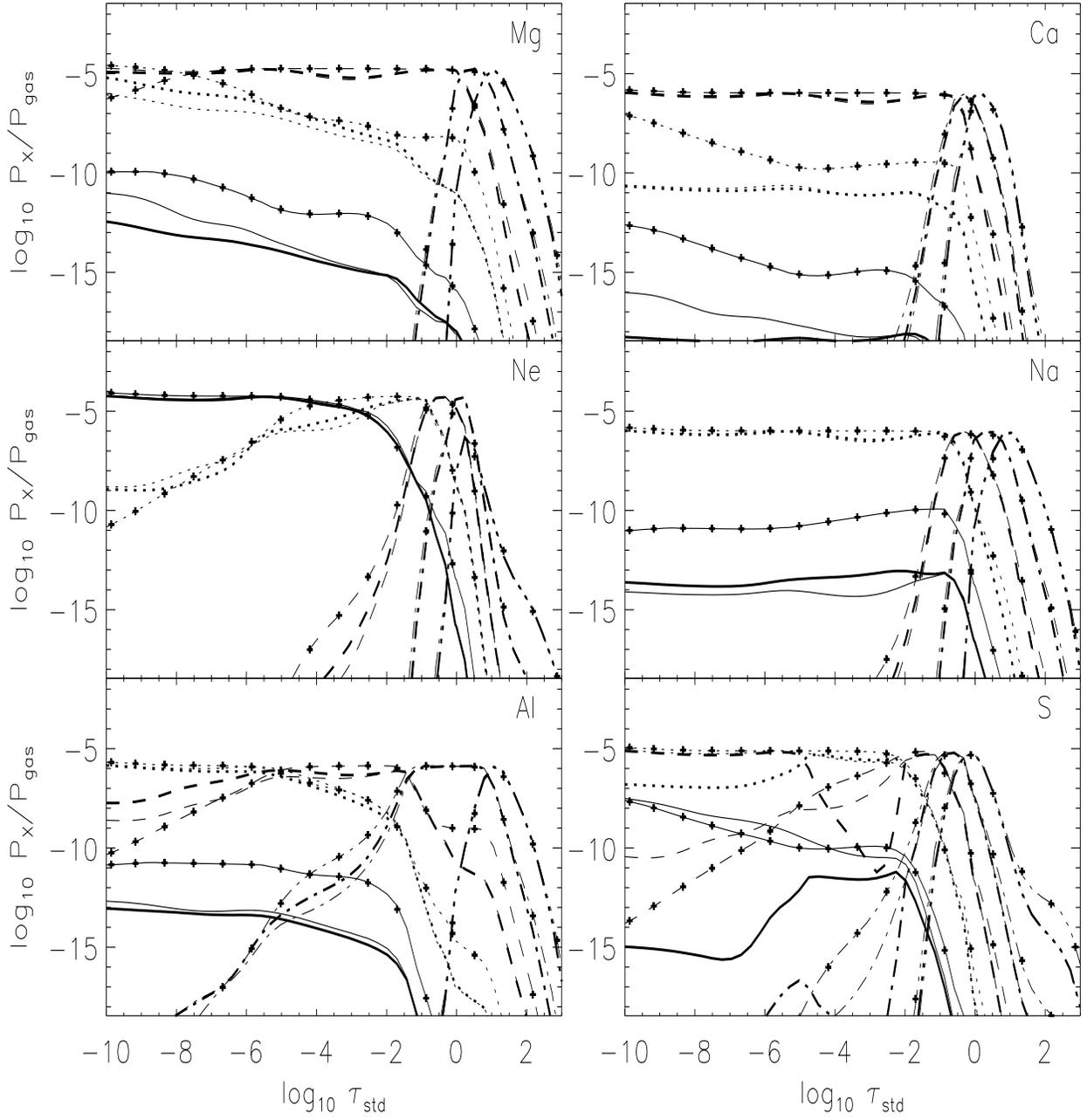}{7.0truein}{0}{100}{135}{-216}{0}
\figcaption[f3.eps]
{Partial pressure of additional species for the \protect $M2$ model.  See Fig. \ref{fpp15}. 
\label{fpp25b} }
\end{figure}

\begin{figure}
\plotfiddle{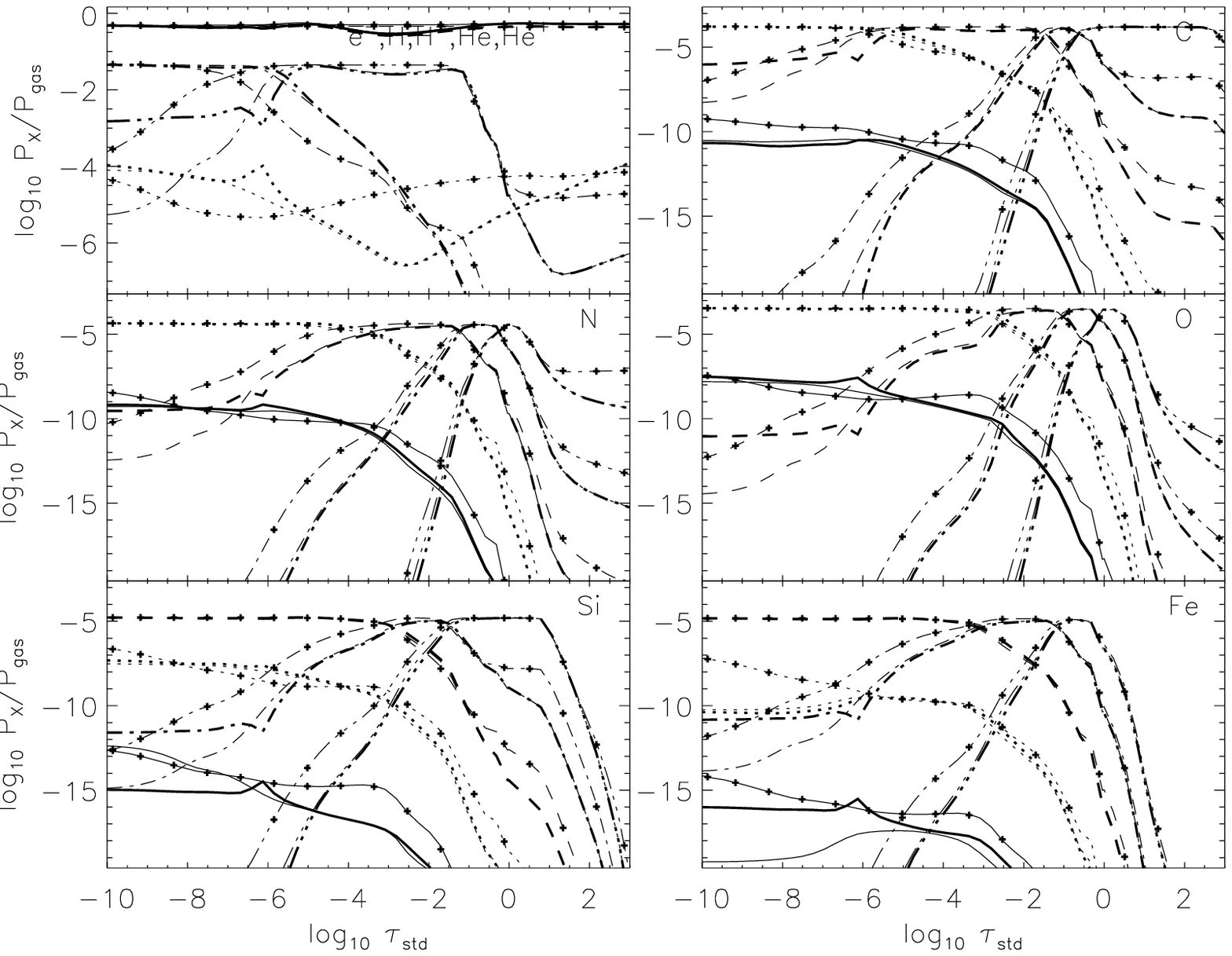}{7.0truein}{0}{100}{135}{-216}{0}
\figcaption[f4.eps]
{Partial pressure of various species for the \protect $M3$ model.  See Fig. \ref{fpp15}. 
\label{fpp35} }
\end{figure}

    Figs. \ref{fpp15} to \ref{fpp35}
show a comparison of the partial pressures of various species computed
in LTE and with both NLTE treatments for all model. 
The current NLTE treatment includes all those species indicated in
Table 3.  Note that 
changes to the NLTE treatment may affect
the concentration of any particular species in three different ways: 
1) through the effect of all NLTE treated species on the chemical 
equilibrium of the 
gas by way of the contribution of each species to the $e^-$ reservoir,
2) through changes in the equilibrium structure of the atmosphere
as a result of NLTE effects on the total opacity and on the $e^-$ density 
(see Fig. \ref{fstruc}), and 3) through changes in the radiation field
in a transition of one species that overlaps an important transition in
another species.  Examples of the latter include line interlocking and
the pumping of transitions, and are especially important for species
that have a rich line spectrum such as Fe, Co, and Ni.

\subsubsection{Hydrogen, Helium, and electrons} \label{srespoph}

The upper left panels of Fig. \ref{fpp15} is directly 
comparable to Fig. 1
of HSSBSA.  We confirm the result of HSSBSA that for a model of $T_{\rm eff}$
equal to $15\,000$ K, NLTE enhances the \ion{H}{1} concentration above the 
LTE values in the 
outer atmosphere where $\log\tau_{\rm cont, 5000}<-6$ and slightly reduces
it in the range where $-6<\log\tau_{\rm cont, 5000}<-2$.  The largest 
NLTE effect is the reduction of the \ion{H}{1} concentration by as 
much as 1.5 dex deep in the atmosphere
where $-1<\log\tau_{\rm cont, 5000}<2.5$.  We note that 
the \ion{H}{2} concentration is very close to the $e^-$ concentration
throughout the atmosphere, which indicates that H ionization is the
dominant $e^-$ contributor, even in the $\log\tau_{\rm cont, 5000}<-4$ range
where H is mostly neutral.  As a result, the NLTE reduction of the $e^-$
concentration in the $\log\tau_{\rm cont, 5000}<-6$ range and enhancement
of it in the $-6<\log\tau_{\rm cont, 5000}<-2$ range is mostly due to
NLTE effects on the H ionization equilibrium.  We also confirm the HSSBSA
result that NLTE slightly reduces the \ion{He}{1} concentration around
$\log\tau_{\rm cont, 5000}=-4$.  We also find that NLTE reduces the
\ion{He}{2} concentration for $\log\tau_{\rm cont, 5000}<-1$.  However,
the \ion{He}{2} concentration is declining rapidly in this range as
$\tau$ decreases due to recombination to \ion{He}{1} as $T(\tau)$ decreases.
This NLTE effect is not apparent in Fig. 1 of HSSBSA because of the
more limited scale of their figure.  The inclusion of the
new species in NLTE has a negligible effect on the concentration of 
\ion{H}{1} and \ion{}{2}, and \ion{He}{1} for this model.  The new treatment 
produces a slight effect on the $e^-$ concentration, first reducing it, then 
enhancing it as $\log\tau_{\rm cont, 5000}$ decreases below $-5$. 
We note that changes in the state of the gas at 
$\log\tau_{\rm cont, 5000}<-6$ only affect the profiles of the strongest
spectral lines. 
The \ion{He}{2} concentration at $\log\tau_{\rm cont, 5000}< -1$
is reduced in the new treatment by approximately as much as the HSSBSA
treatment reduced it from the LTE values.  
%However, when considering these 
%deviations in the new results, we note that the model \ion{H}{1} atom 
%employed here has 80 levels, while that of HSSBSA has 30.  The enlargement
%of the \ion{H}{1} atom affects the \ion{H}{1}/\ion{}{2} equilibrium, and, 
%therefore, the $e^-$ concentration and $P_{\rm gas}$ by providing 
%additional recombination channels. 

\paragraph{} 

  The upper left panels of Fig. \ref{fpp25a} shows the same 
species for the $M2$ model.  Here, the 
effects of NLTE, and of the different NLTE treatments are larger. 
For $\log\tau_{\rm cont, 5000}< -4$ where \ion{H}{1} becomes relatively
abundant, the HSSBSA treatment reduces the \ion{H}{1} concentration by 
as much as two and a half orders of magnitude.  However, the current
NLTE treatment gives rise to a {\it smaller} reduction, producing results
that are closer to the original LTE concentration.  The new treatment also
predicts a small enhancement in the concentration around 
$\log\tau_{\rm cont, 5000}=-5$ that is not present in the HSSBSA results.  
As with the $M1$ model, the largest NLTE effect is reduction in \ion{H}{1}
concentration by as much as $\approx 1.5$ dex at greater depth where
$-3<\log\tau_{\rm cont, 5000}<2.5$.
Both NLTE treatments give rise to very similar He concentrations and
produce a reduction in the \ion{He}{1} and \ion{}{2} pressures for 
$\log\tau_{\rm cont, 5000}< -1.5$.  This reduction is as much as half an 
order of magnitude in the case of \ion{He}{1}.  Unlike the $M1$ model, 
the $e^-$ pressure throughout the outer atmosphere is {\it not} driven by 
either the H or He ionization equilibria.  The two regions of
reduction in the $e^-$ concentration at $-5<\log\tau_{\rm cont, 5000}<-1$
and $-9<\log\tau_{\rm cont, 5000}<-6$ are matched by similar {\it reductions}
in \ion{He}{1} concentration and the outermost of these reductions is also
matched by a {\it reduction} in the \ion{H}{1} concentration.  A NLTE 
reduction in
the population of electrons available for recombinations is driving a 
reduction in the neutral H and He populations.

\paragraph{}

   Model $M3$ is hotter than any of the models discussed in HSSBSA.  From the 
upper left panel of Fig. \ref{fpp35} we see that at $T_{\rm eff}=35\,000$ K
H is almost completely ionized throughout the entire atmosphere.  Neutral H 
is a very small minority stage and is very sensitive to the treatment of the
equilibrium.  The NLTE suppression of \ion{H}{1} at depth that was noted
for the $M1$ and $M2$ models is more pronounced here: the NLTE \ion{H}{1} 
concentration drops below its LTE value
by as much as two orders of magnitude from $-5<\log\tau_{\rm cont, 5000}<2.5$.  The
$e^-$ and H$^+$ populations, which dominate the gas pressure throughout the 
entire atmosphere, show slight NLTE effects around 
$-4<\log\tau_{\rm cont, 5000}<-2$ where there is a slight NLTE reduction by
as much as $\approx 0.3$ dex. 

\subsubsection{Metals} \label{srespopm}

   Carbon, nitrogen, and oxygen (CNO) are important in nova spectra as a 
measure of convective dredge up 
of nuclear processed material in the progenitor WD, and
as a test of thermonuclear runaway (TNR) models of nova 
explosions (\cite{starr1}, \cite{gehrz}).
The upper right panel of Fig. \ref{fpp25a} is directly comparable
to Fig. 2 of HSSBSA.  We confirm their result that \ion{C}{2} is the
dominant ionization stage, and that the HSSBSA NLTE
treatment reduces the \ion{C}{1} and enhances the \ion{C}{3} concentration 
compared to the LTE levels for $\log\tau_{\rm cont, 5000}<-2$.
The current NLTE treatment produces approximately the same results for
\ion{C}{1} and \ion{C}{2}, but produces a slight NLTE {\it reduction}
in \ion{C}{3} throughout most of the range where the HSSBSA treatment
produces an enhancement.  From examination of Fig.
\ref{fpp25a} and of Fig. 3 in HSSBSA we see that the same relative behavior 
for the different NLTE treatments is seen for the lowest three stages of N.
From inspection of Fig. \ref{fpp15} and 
Fig. 4 in HSSBSA we see
that NLTE effects for O are much smaller in the cooler $M1$ model.  We note that
the details of the NLTE treatment of CNO are identical in for both HSSBSA
and the present calculation; the number of ionization stages and the size
of the model atoms are the same in both treatments.  The large differences
seen in the concentration of some CNO stages in the $M2$ model are entirely
due to the {\it indirect} effects of other species treated in NLTE by way
of either the chemical equilibrium or the atmospheric structure.  

\paragraph{}

  The rich blanketing of line opacity
contributed by Fe plays a crucial role in the time development of 
the $UV$ and optical light curves and spectra during the outburst 
(\cite{hausfe}, \cite{shore}).  From inspection of the lower right panel in
Figs. \ref{fpp15} and \ref{fpp25a} we see that \ion{Fe}{2} is the dominant
ionization stage throughout most of the outer atmosphere in the $M1$ model,
whereas \ion{Fe}{3} plays the same role in the $M2$ and $M3$ models.  For the
$M1$ model, the HSSBSA NLTE treatment leads to a significant enhancement
of the \ion{Fe}{1} concentration above LTE values for 
$\log\tau_{\rm cont, 5000}<-2$, 
whereas the current treatment leads to a {\it reduction} in this range and
to a concentration that is much closer to LTE.  Both treatments lead to 
significant enhancements in the \ion{Fe}{3} concentration throughout the 
outer atmosphere.  For the $M2$ and $M3$ models, NLTE effects in both treatments lead
to a large reduction in both the \ion{Fe}{1} and \ion{Fe}{2} concentrations
for $\log\tau_{\rm cont, 5000}<-2$.  However, the current NLTE treatment gives
rise to an \ion{Fe}{1} reduction that is approximately twice as large as the
that of the HSSBSA treatment.  We note that \ion{Fe}{2} in particular has
a very rich $UV$ line spectrum that determines the $UV$ flux distribution and
affects the atmospheric structure during the optically thick wind phase
of the outburst (\cite{hausfe}, \cite{shore}).  Therefore,
the reduction in \ion{Fe}{2} concentration by two orders of magnitude that 
is produced by both NLTE treatments in the $M2$ and $M3$ models will have a 
significant effect on the model structure.  This is also discussed by HSSBSA. 
%Unfortunately, HSSBSA do not show Fe concentrations
%for a model of the temperature of the $M1$ or $M2$ models, so we can not
%directly compare our results to theirs.  
As was the case with CNO, differences
between the two NLTE treatments may be due to the indirect effects of other
NLTE species.  In addition, the difference in the \ion{Fe}{1}-\ion{}{3}
results may also be due to the presence of a detailed \ion{Fe}{4} model
atom in the NLTE calculation. 
 
\paragraph{}

   In Figs. \ref{fpp25a} and \ref{fpp25b}
we show concentrations for a variety of other species treated in NLTE for the
$M2$ model.  NLTE
results for S and Si were shown in HSSBSA, but they only treated the second and
third ionization stages in NLTE, whereas we treat the lowest six stages.
These two elements provide an object lesson in the importance of adjacent 
ionization stages to the accurate treatment of a particular stage in NLTE.
In their Fig. 5 HSSBSA show an increase in the concentration of
\ion{S}{1} with respect to LTE by as much as three orders of magnitude 
for $\log\tau<-3$ when \ion{S}{2} and \ion{}{3} are treated in NLTE.  
Our results for the $M2$ model,
shown in the bottom right panel of Fig. \ref{fpp25b}, are not exactly
comparable because $M2$ has a $T_{\rm eff}$ that is $5000$ K cooler
than the model for which HSSBSA show results.  However, 
our treatment, in which \ion{S}{1} is included in NLTE, produces a seven 
orders of magnitude
{\it decrease} in the \ion{S}{1} concentration in the outer atmosphere.  
Moreover, whereas, HSSBSA found an increase by $\approx 1$ order of 
magnitude 
in the \ion{S}{3} concentration in the same $\tau$ range, our
calculations, in which \ion{S}{4} is included in NLTE, show an increase 
by as much as nine
orders of magnitude in the outer atmosphere.  Comparison of Fig. 6 in 
HSSBSA to the bottom left
panel of Fig. \ref{fpp25a} shows similar effects for Si.  For example, 
whereas the HSSBSA treatment, in which \ion{Si}{1} is in LTE, yields an 
increase with respect to LTE in the \ion{Si}{1} concentration in the outer
atmosphere of a $25\, 000$ K model, we find that the $M2$ model with
\ion{Si}{1} to \ion{}{4} in NLTE shows a {\it decrease} by as much as 
five orders of magnitude.
We note that \ion{S}{1} and \ion{Si}{1} are both minority stages in these
models.  In general, the population of a minority stage is very sensitive
to the photoionization rate and to the population of the reservoir stage.

\paragraph{} 

Aluminum and the higher stages of Na are treated in NLTE for the first time.
Magnesium and Ca both have $UV$ resonance lines from the second ionization
stage that are important spectral features.  HSSBSA treated only the second
stage in NLTE, whereas we treat the lowest six.  Inspection of these figures
shows that for all of these species, NLTE effects, and the particular treatment
of NLTE, are significant for one or more stages.

\subsection{Atmospheric structure} \label{sresstruc}

\begin{figure}
\plotone{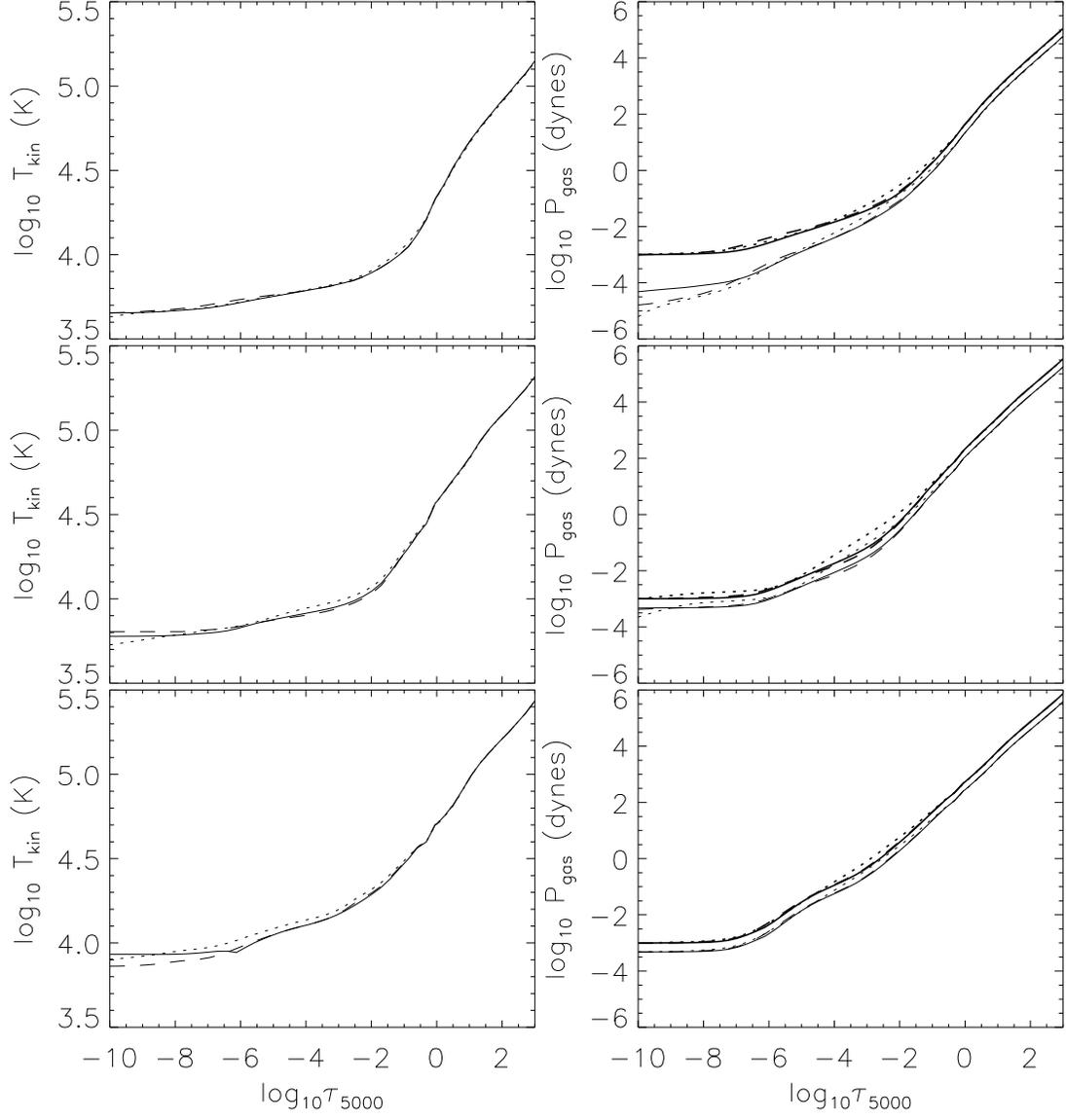}
\figcaption[f5.eps]
{Atmospheric structure of models.  Left panels: kinetic temperature, right 
panels: gas and \protect $e^-$ pressure.  Upper panels: \protect $M1$ model, middle panels: 
\protect $M2$ model, lower panels: \protect $M3$ model.  
Solid line: current NLTE, dashed line: HSSBSA NLTE, dotted line: LTE.  
In the right panels: thin lines:
\protect $e^-$ pressure, thick lines: total gas pressure. \label{fstruc} }
\end{figure}

   Fig. \ref{fstruc} shows the atmospheric structure of our models.  From
the left panels we see that the HSSBSA NLTE treatment gives rise to
surface heating with respect to LTE for $\log\tau_{\rm cont, 5000}<-6$ in the $M2$ model.  
By contrast, the current NLTE treatment gives rise to a $T_{\rm kin}$ structure
that is cooler in the upper atmosphere.
The HSSBSA treatment yields a surface 
cooling with respect to LTE in the $M3$ model whereas the more complete NLTE treatment produces 
higher $T_{\rm kin}$ values in the upper atmosphere.  For both models, the more
complete NLTE treatment gives rise to a $T_{\rm kin}$ structure that is {\it closer}
to the LTE structure than that of the less complete NLTE treatment.
  We note that the presence or absence of this NLTE
surface heating or cooling in the models will affect the cores of strong lines 
that form 
near the top of the atmosphere.  NLTE effects on the $T_{\rm kin}$ structure are
negligible in the $M1$ model.
From the right panels of Fig. \ref{fstruc} we see that NLTE effects
cause a slight reduction in $P_{\rm gas}$ around 
$\log\tau_{\rm cont, 5000}=-2$ in the $M1$ model and a general reduction
throughout the atmosphere in the $M2$ model for $\log\tau_{\rm cont, 5000}<-2$.  
Both NLTE treatments give approximately the same result.  For the $M3$ model, 
both NLTE treatments give pressure structures that are close to the LTE structure 
except for a slight reduction 
around $-4<\log\tau_{\rm cont, 5000}<-2$. 
By comparing the $P_{\rm gas}$ and $P_{\rm e}$
structure, we note that the NLTE deviations in $P_{\rm gas}$ in all models 
mirror those
in $P_{\rm e}$.  The NLTE behavior of $P_{\rm e}$ was discussed above in 
connection with NLTE H and He.

\subsection{Flux distribution} \label{sresflux}

\begin{figure}
\plotone{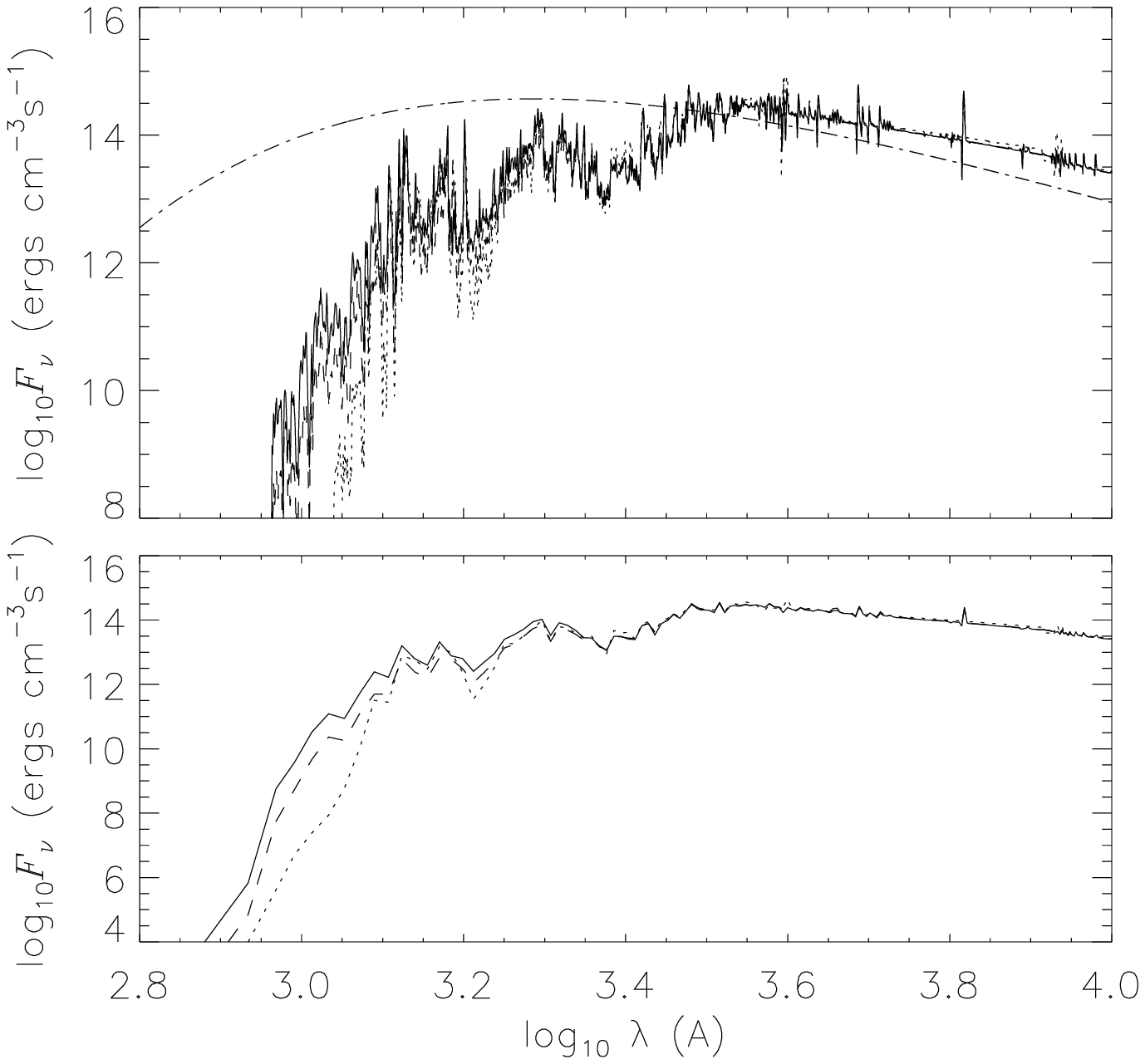}
\figcaption[f6.eps]
{Flux distribution, \protect $F_\lambda$, of \protect $M1$ model. 
Upper panels: flux distribution computed with \protect $\Delta\lambda=1.0$\AA~ for 
\protect $\lambda<900$\AA~ and with \protect $\Delta\lambda=0.5$\AA~ for \protect $\lambda>900$\AA.  
Lower panels: flux distribution smoothed with a boxcar of \protect $\Delta\lambda=100$\AA~ 
in the far \protect $UV$ and \protect $\Delta\lambda=50$\AA~ in the near \protect $UV$, optical, and \protect $IR$.  
Solid line: current NLTE, dashed line: HSSBSA NLTE, dotted line: LTE,  
dot-dashed line: Planck function (\protect $B_\lambda$). \label{ffluxm1} } 
\end{figure}

\begin{figure}
\plotone{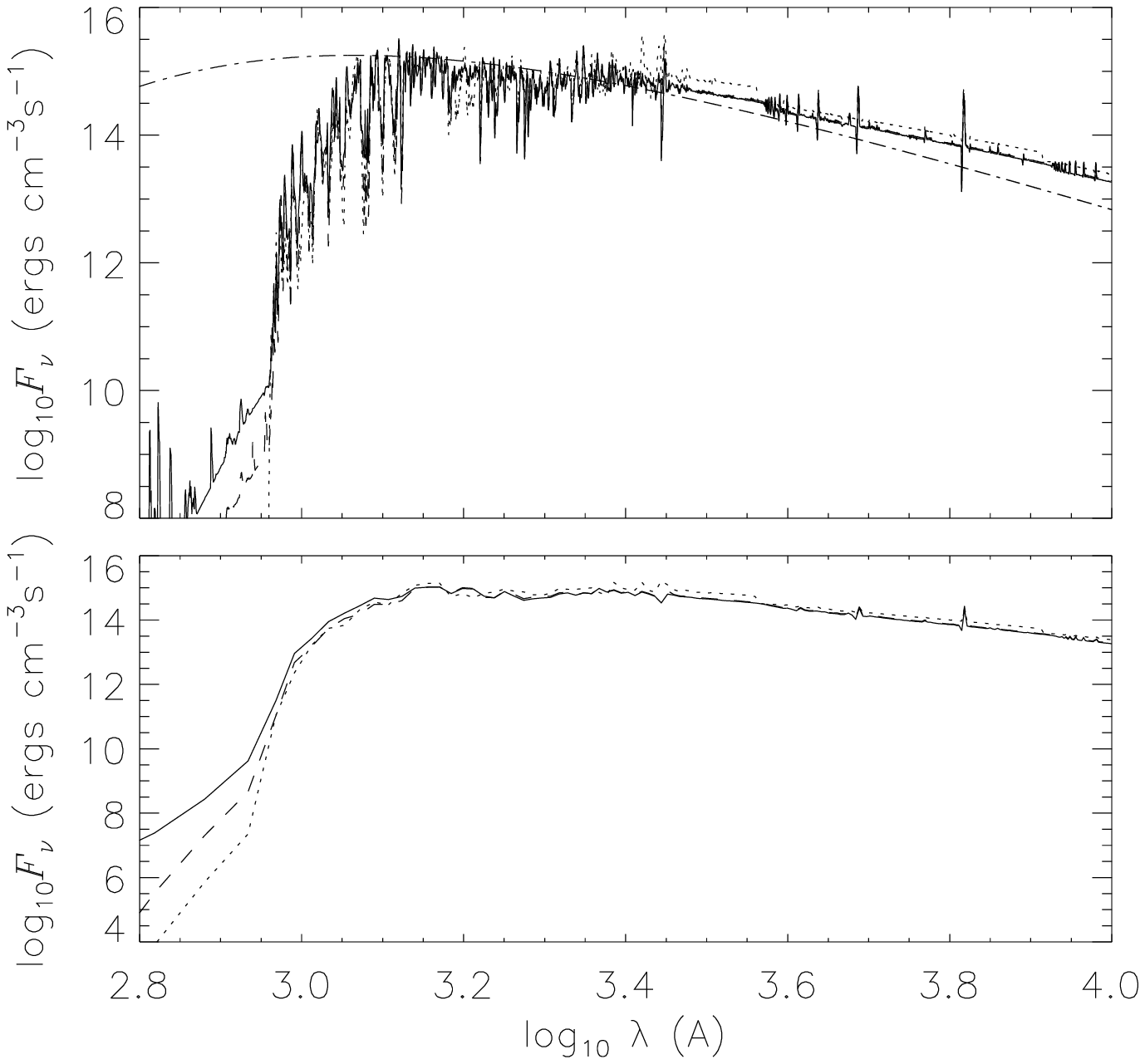}
\figcaption[f7.eps]
{Flux distribution, \protect $F_\lambda$, of \protect $M2$ model.  See Fig. \ref{ffluxm1}. 
\label{ffluxm2} } 
\end{figure}

\begin{figure}
\plotone{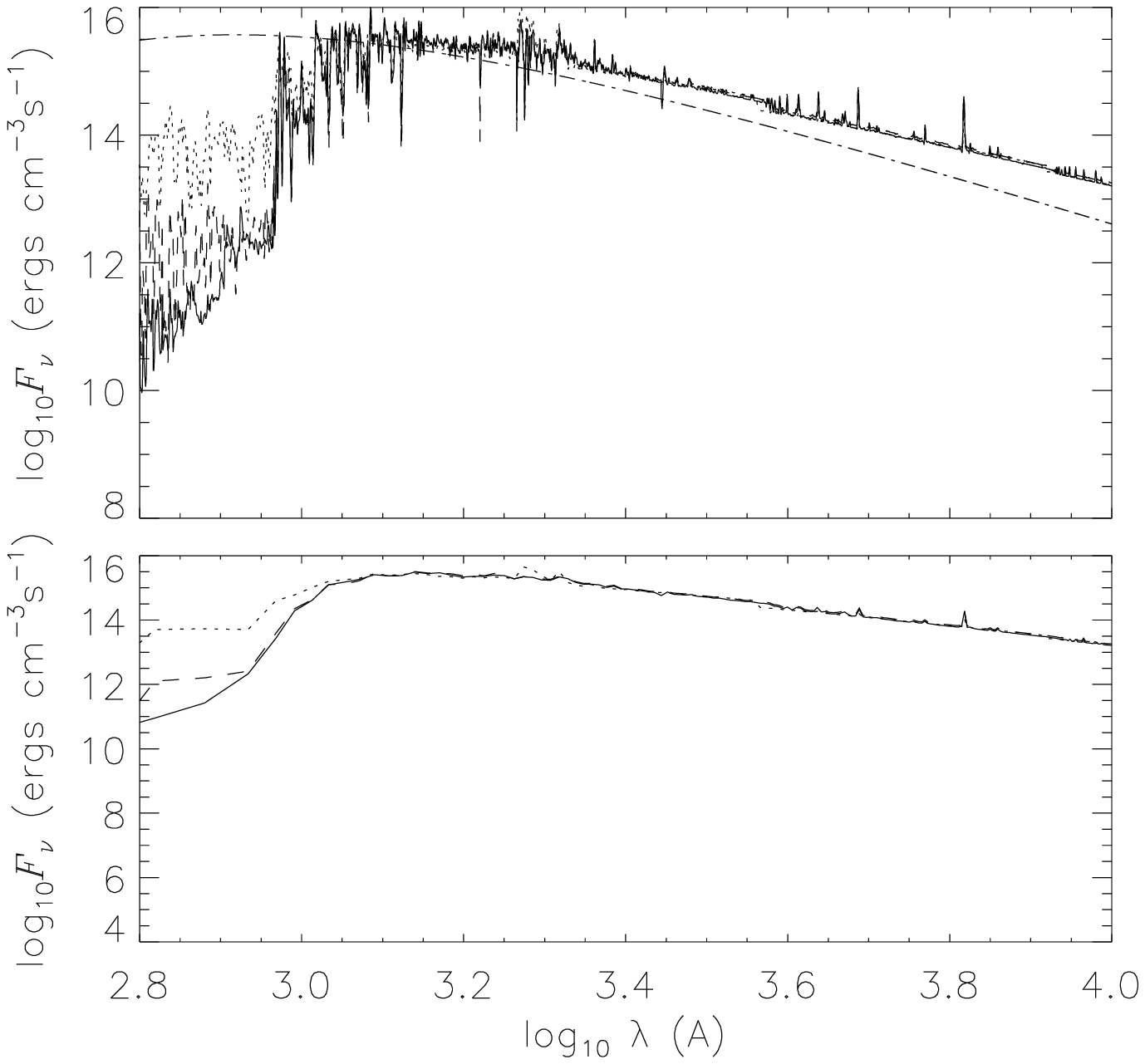}
\figcaption[f8.eps]
{Flux distribution, \protect $F_\lambda$, of \protect $M3$ model.  See Fig. \ref{ffluxm1}. 
\label{ffluxm3} } 
\end{figure}

   Figs. \ref{ffluxm1} to \ref{ffluxm3} show the distribution of the emergent flux, 
$F_\lambda(\tau=0)$ produced by the models in the observer's (Eulerian) frame.  In 
the upper panel we show
$F_\lambda$ at the computed resolution of $\Delta\lambda=1.0$\AA~ for 
$\lambda<900$\AA, and $\Delta\lambda=0.5$\AA~ for $\lambda>900$\AA.
The effect of massive line blanketing on the spectrum can be seen for
$\log\lambda<3.5$.  In the lower panel we show the same $F_\lambda$ distributions
after smoothing with a boxcar function of width $\Delta\lambda=100$\AA~ in the 
far $UV$ and $\Delta\lambda=50$\AA~ in the near $UV$, optical and near infrared ($IR$).  The
smoothing approximately reproduces the resolution of intermediate band
photometry and wide band spectrophotometry and allows differences in the
overall $F_\lambda$ level to be more easily discerned.  

\paragraph{}

From the lower panels
we see that NLTE effects lead to large enhancements in $F_\lambda$ for 
$\log\lambda<3.1$ in the $M1$ model and for $\log\lambda<2.95$ in the $M2$ model.  
By contrast, NLTE effects
lead to a {\em reduction} in the UV flux in the $M3$ model for $\log\lambda<2.95$. 
For all three models, the
current NLTE treatment increases the size of the NLTE deviation in
$F_\lambda$ significantly.  
The opacity sources that have the largest effect on the emergent flux below
the Lyman edge ($\log\lambda<2.95$) are
\ion{H}{1} $b-f$ absorption and line absorption.  We see from Figs. 
\ref{fpp15} through \ref{fpp35} that the \ion{H}{1} concentration
is always significantly reduced by the effect of NLTE at depths
where $\tau_{\rm cont, 5000}\le 1$, and 
is often enhanced higher up in the atmosphere.  Furthermore, the current
NLTE treatment serves variously to enhance or diminish the size of the NLTE
deviation in \ion{H}{1} concentration.  We may expect that, all
else being equal, models in which the emergent flux in the Lyman continuum 
arises from deeper in the atmosphere will be brighter in the UV in the case
of NLTE because the reduction in \ion{H}{1} will allow flux to escape from
deeper, hotter layers.  By contrast, models in which this flux arises from higher 
up in the atmosphere will be dimmer in the case of NLTE.  The extent of the 
NLTE effect on the emergent flux for particular NLTE treatments will by modified 
by how that treatment effects the \ion{H}{1} concentration.

\paragraph{}

    The situation is complicated by the competing effect of line opacity in
determining the emergent UV flux.  The net effect of NLTE departures on all 
the metals that contribute significant line opacity may serve either to
brighten or dim the flux.  From an examination of the concentrations of the
dominant Fe ionization stages in Figs. \ref{fpp15} through \ref{fpp35}, we see that in 
the case of $M1$ the amount of Fe line blanketing is enhanced by the current
NLTE treatment compared to the other two treatments, and in $M2$ and $M3$ 
Fe line blanketing is reduced by both NLTE treatments.  It is not clear
how this can be reconciled with the effects of NLTE on the flux seen in Figs. 
\ref{ffluxm1} through \ref{ffluxm3}, and we conclude that the competing influence
of NLTE on all the various absorbers that contribute significant opacity is 
too complex
for a simple correlation between the emergent flux and the concentration of any
one absorber to be apparent.  Finally, we note that generally the flux on the
Wien side of the Planck distribution is a sensitive 
indicator of $T_{\rm eff}$.  Therefore, the significant NLTE effects on the UV
flux seen here must be included accurately in models.

\subsection{Completeness} \label{srescomp}

\begin{figure}
\plotone{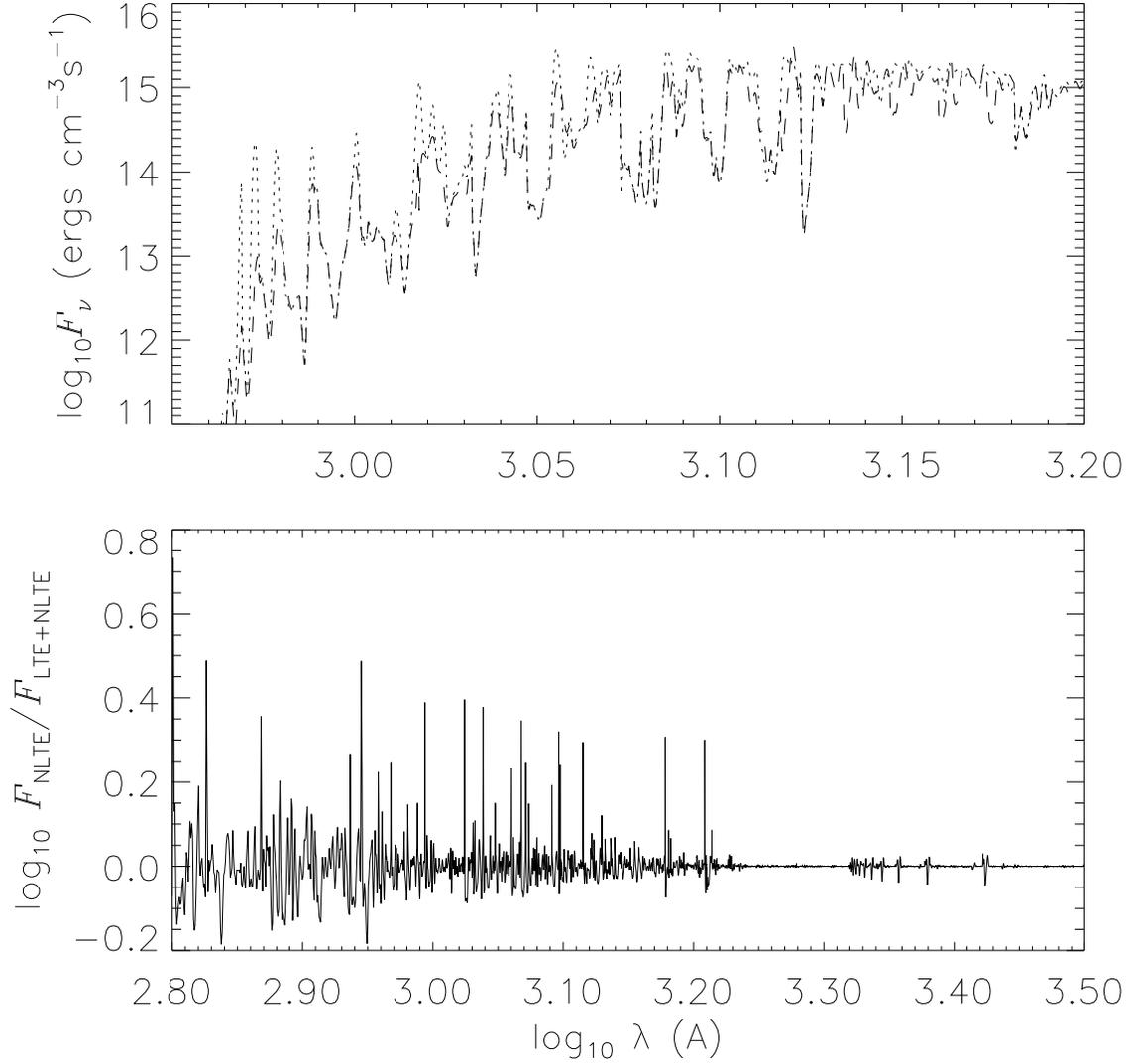}
\figcaption[f9.eps]
{Flux distribution, \protect $F_\lambda$, of \protect $M2$ model with
different treatments of line blanketing.  Upper panel: \protect $F_\lambda$ with
NLTE lines only: dotted line; \protect $F_\lambda$ with LTE and NLTE lines (complete
blanketing): dashed line.
Lower panel: Ratio of \protect $F_\lambda$ computed with NLTE lines only to \protect $F_\lambda$
computed with all lines.
\label{figcomp} } 
\end{figure}
   
Fig. \ref{figcomp} shows the overall $F_\lambda$ distribution for the $M2$
model with complete line blanketing due to all species that contribute 
significant line opacity (whether they are treated in LTE or NLTE),
and with line blanketing due {\it only} to those species treated in NLTE,
as indicated in Table 3.  A comparison of 
the two distributions allows an assessment of the completeness of
the line opacity that is now treated in NLTE.  This is important
for assessing the accuracy of the emergent flux spectrum, and also
of the equilibrium atmospheric structure because line opacity is
an important term in the radiative equilibrium of the atmosphere.  

\paragraph{}

The most obvious difference between the two $F_\lambda$ distributions
is that the Lyman photo-ionization edge at $\log\lambda=2.96$ is ``softened''
by line blanketing to a greater extent when LTE lines are present. 
The missing NLTE UV line opacity will mostly be accounted for 
once Fe group species that have a rich line spectrum,
such as \ion{Cr}{3} and \ion{Ni}{2} and \ion{}{3} and \ion{Co}{2}
have been added in NLTE. 
From Table 1 it can be seen that we already have the facility to
compute Ni in NLTE up to stage \ion{}{6} and Co up to stage \ion{}{3}, 
but, as noted in Section \ref{smods} and in Table 3,
these stages have not been included in these calculations for the
sake of computational expediency. 
Also, there is a noticeable underblanketing in the 
$3.1<\log\lambda<3.5$ range in the treatment with NLTE lines only compared
to the fully blanketed treatment.  Nevertheless, from the qualitative 
similarity of the two $F_\lambda$ distributions we can also see that 
the lines treated in NLTE account for the vast majority of the total
line opacity.  

\subsection{$UV$ and optical spectra} \label{sresspec}  

\begin{figure}
\plotone{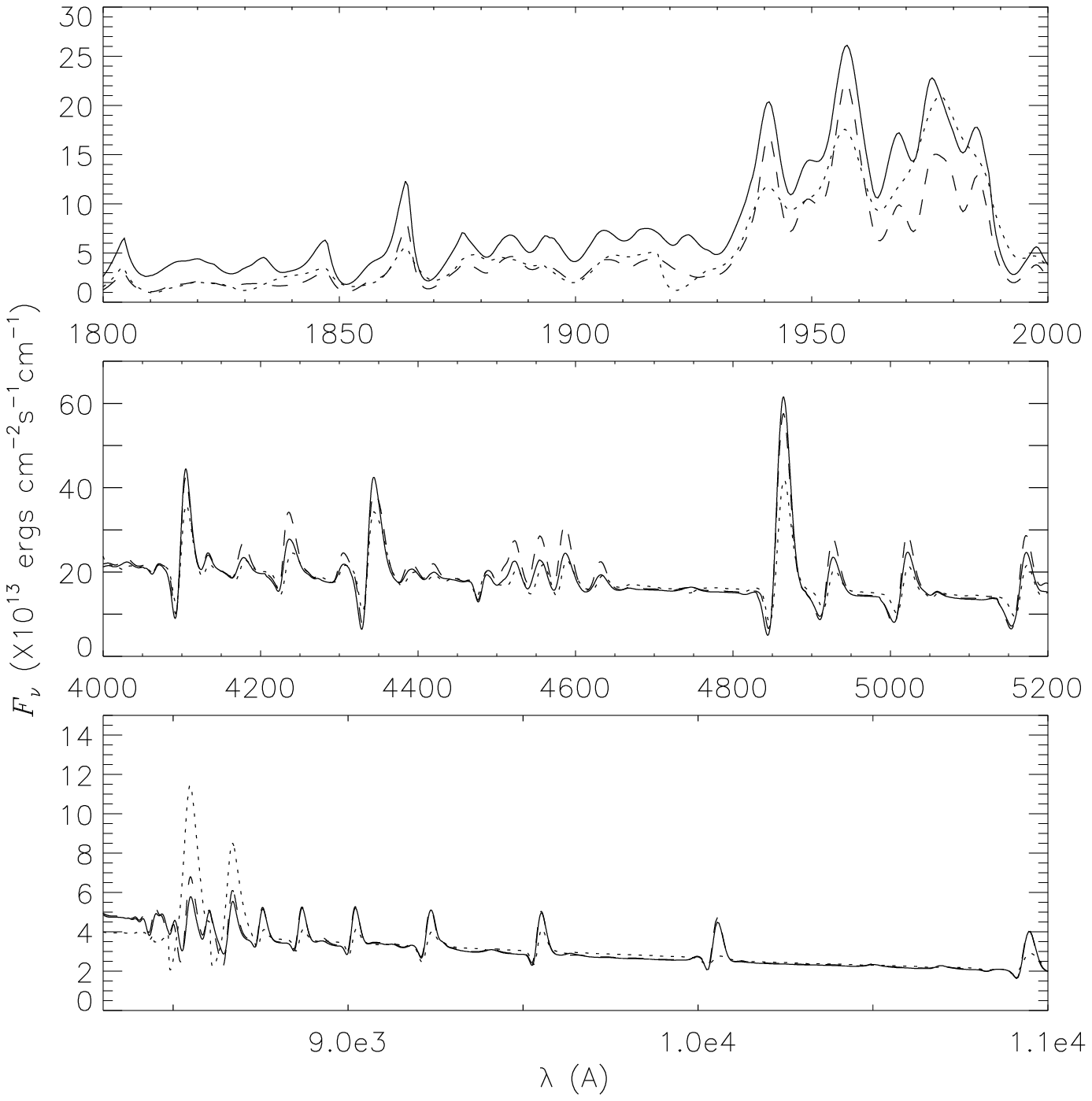}
\figcaption[f10.eps]
{Flux spectrum of \protect $M1$ model in three sample regions from 
the mid-\protect $UV$ to near \protect $IR$.  
Solid line: current NLTE, dashed line: HSSBSA NLTE, dotted line: LTE,  
\label{fspec15} } 
\end{figure}

\begin{figure}
\plotone{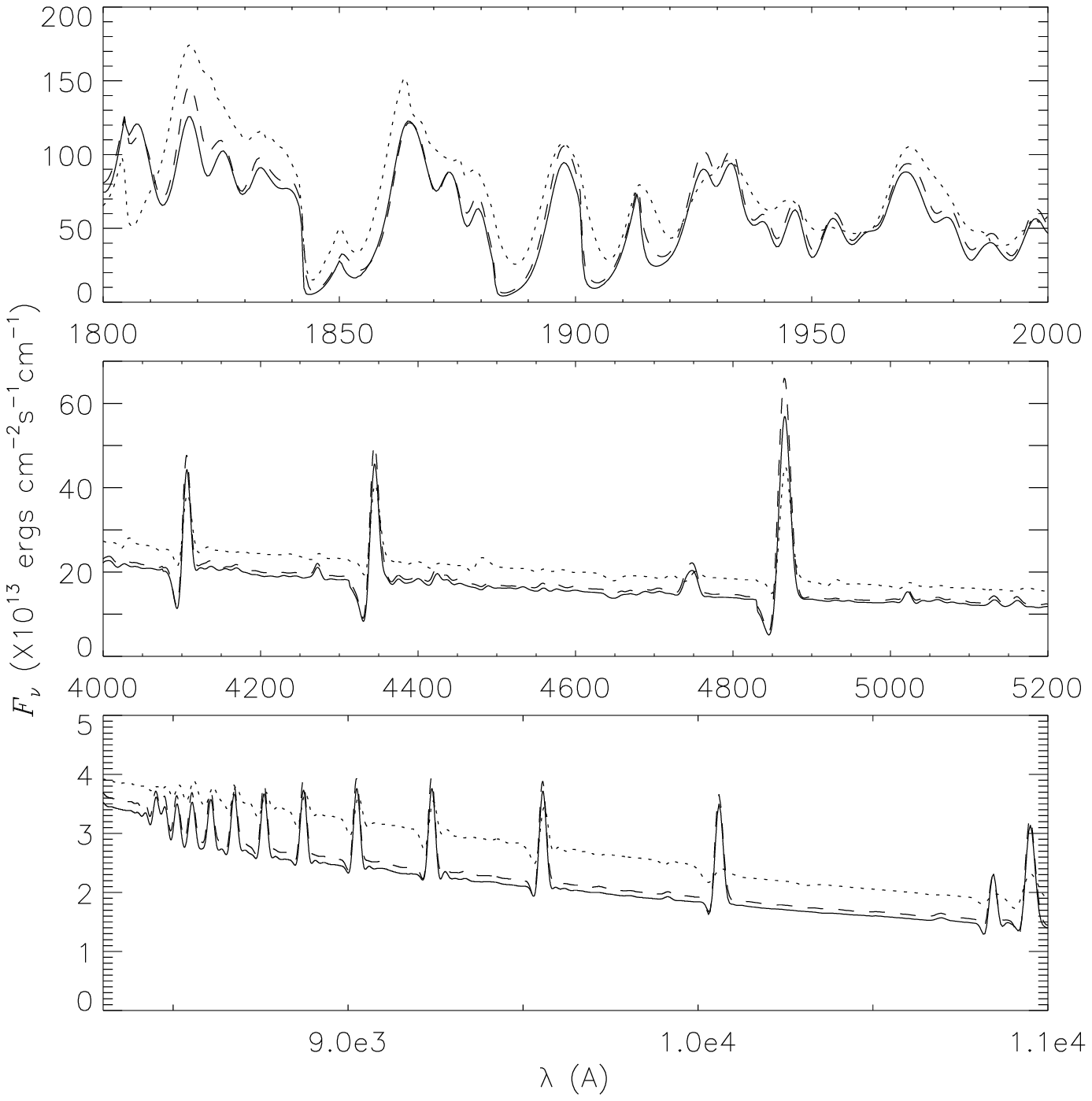}
\figcaption[f11.eps]
{Flux spectrum of \protect $M2$ model in three sample regions from mid-\protect $UV$ to near \protect $IR$. 
See Fig. \ref{fspec15} \label{fspec25} } 
\end{figure}

\begin{figure}
\plotone{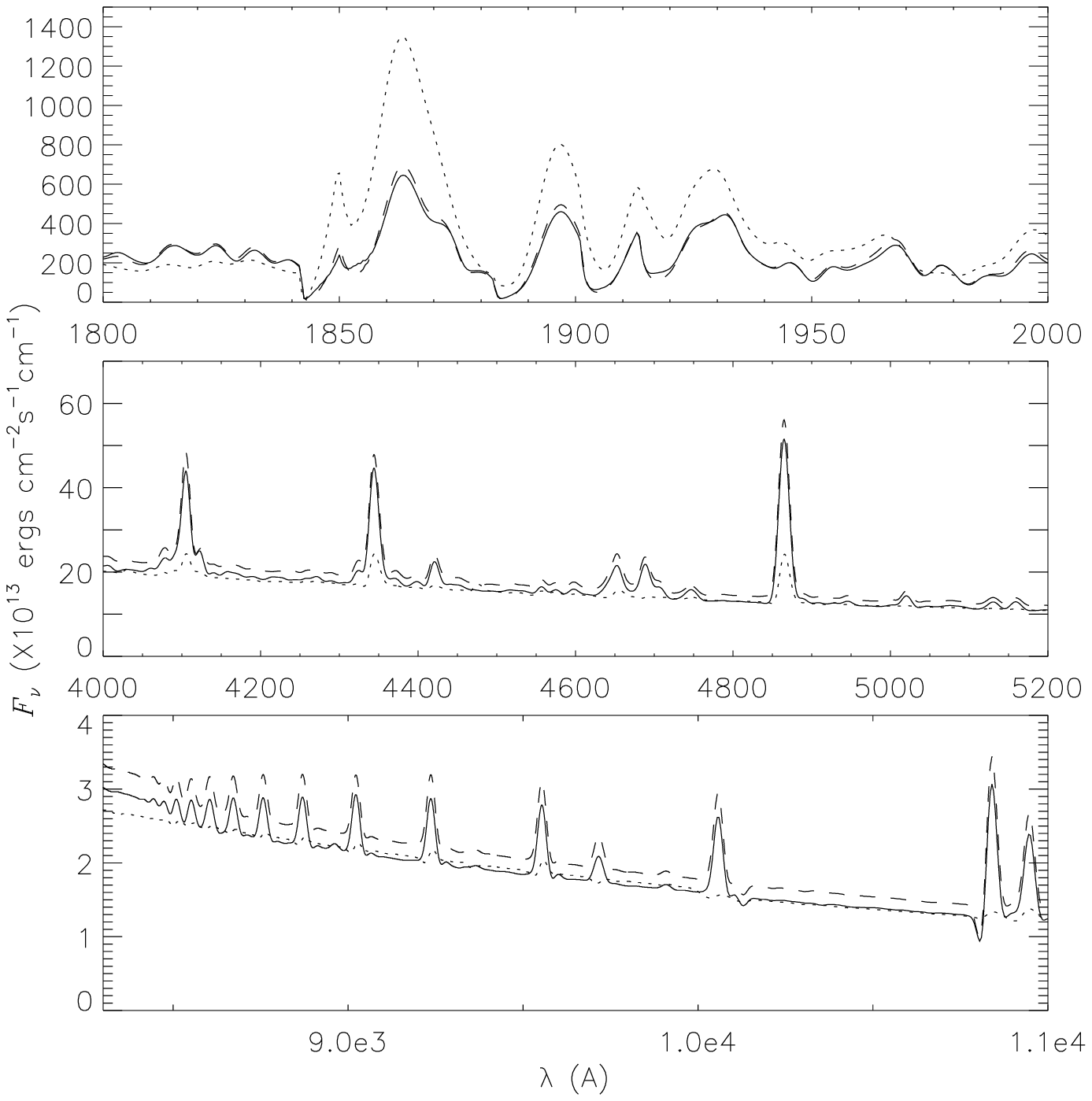}
\figcaption[f12.eps]
{Flux spectrum of \protect $M3$ model in three sample regions from mid-\protect $UV$ to near \protect $IR$. 
See Fig. \ref{fspec15} \label{fspec35} } 
\end{figure}

   Figs. \ref{fspec15} to \ref{fspec35} show the moderate resolution spectrum
in three sample regions from the mid-$UV$ to the near-IR. 
Like HSSBSA we find that 
the quasi-LTE spectrum with a single scattering albedo incorporated into
the line source function gives rise to lines that are approximately equal in 
strength to NLTE lines.  The quasi-LTE approach gives rise to inaccurate 
line strengths for many particular lines, but does not systematically 
over-predict or under-predict line absorption.  Inspection of Figs. 
\ref{fspec15} and \ref{fspec25} shows that the HSSBA and the current NLTE 
treatment give rise to different line profiles for many lines, but the differences
are not systematic.  Differences between the two NLTE treatments are particularly  
apparent in the $UV$ region of the $M2$ model as shown in the top panel of 
Fig. \ref{fspec25}.  Note that differences in 
the profiles of individual lines that arise from the NLTE treatment 
may in general be due to a combination of
effects: 1) changes in the line source function, $S_\nu(\tau)$, as a result
of changes in the populations, $n_{\rm l}$ and $n_{\rm u}$, of the levels
connected by the transition and by the inclusion of a properly calculated 
scattering contribution in $S_\nu(\tau)$, 2) changes in the background 
continuous and line opacity due to NLTE effects in other species, and
3) changes in the atmospheric structure such as those discussed in Section
\ref{sresstruc}.  In general, NLTE effects should be included in as complete
a way as possible to accurately calculate any particular line profile.

\section{Conclusions} \label{sconc}

    Generally, we find that the current, more complete NLTE calculation gives 
rise to {\it qualitatively} the same results as the previous extensive NLTE
investigation of nova models, that of HSSBSA.  NLTE may greatly affect:
1) the chemical concentration of
species that are important to the total opacity of the model, such as \ion{H}{1} and Fe,
or are astrophysically interesting, such as CNO, Mg, Al, and Ca, 2) the overall flux 
distribution, particularly in the $UV$ ($\lambda<1000$\AA), which is a sensitive 
 $T_{\rm eff}$ indicator, and  
where NLTE effects increase the flux by several orders of magnitude, 3)
the strength of individual lines throughout the observable $UV$ to $IR$ region
where NLTE causes particular lines to be either weaker or stronger than
those of a quasi-LTE calculation that crudely incorporates NLTE effects with
a single scattering albedo, and 4) to a lesser extent, 
the atmospheric structure, particularly the $P_{\rm gas}$ structure in models 
of $T_{\rm eff}\approx 25\, 000$ K.  However, the current NLTE treatment
gives rise to specific deviations from LTE values for many of these quantities
that differ significantly from those of the HSSBSA treatment.  In particular,
we note that for the hotter $M2$ model ($T_{\rm eff}=25\, 000$ K), in which
NLTE effects are generally larger than for the cooler $M1$ model 
($T_{\rm eff}=15\, 000$ K), the more complete NLTE treatment gives rise to
an \ion{H}{1} concentration, $T_{\rm kin}$ and $P_{\rm gas}$ 
structure, and $UV$ flux level that are all significantly {\it closer} 
to the LTE values than those of the less complete HSSBSA NLTE treatment. 
This result, which may seem counterintuitive at first, results from the 
increased number of channels through which the gas can thermalize when
more species are included in the overall NLTE solution.  Note that
increasing the number of NLTE species does not necessarily drive the
solution for any model atmosphere closer to LTE.  Rather, for the $M2$
model in particular, the increase in the number of thermalization channels
happens to have a dominant effect on the solution when the number of 
NLTE species is increased from the HSSBSA set to the current set.       

\paragraph{}

  In general, NLTE should be incorporated in as complete a way as possible
for accurate structure and synthetic spectrum calculations for nova
models in the 15\, 000 to 35\, 000 K $T_{\rm eff}$ range.  In regard to the
first point above, we draw special attention to the {\it indirect} effects
that treating any one species in NLTE may have on any other species.  The
atmospheric structure and the $e^-$ concentration both act as means of
coupling all the species, including those that are treated in LTE.  Generally,
all species that contribute significantly to the $e^-$ reservoir by way of
partial ionization, or whose
line or continuous opacity is large enough to significantly effect the 
equilibrium structure of the atmosphere, must be treated in NLTE to insure
an accurate result for any particular species.  This has a bearing
on, for example, attempts to infer model parameters and abundances from
fitting the $UV$ and optical light curves of novae during the optically thick
wind phase of their outburst. 

% Authors may indicate to the editorial staff where they would like 
% figures and tables to be placed in the manuscript.  This is done with
% either the \placefigure{KEY} or \placetable{KEY} commands.  These
% commands require \label{KEY} commands to be placed appropriately with
% corresponding table and figure captions.  When the manuscript is
% printed a short note is printed on the page where the figure or table
% is to go.  These commands are ignored in the aaspp4 and aas2pp4 styles.

% The \notetoeditor{TEXT} command allows the author to communicate some
% information to the copy editor.  This information will appear as a 
% footnote on the printed copy for the aasms4 style file.  Nothing will 
% appear on the printed copy if the aaspp4 or aas2pp4 style file is used.

% This is the last section of the paper, so there is an \acknowledgments
% section at the end of the main body.

\acknowledgments

This work was supported in part by NASA ATP grant NAG 5-3018 and LTSA grant NAG
5-3619 and NSF grant AST-9720804 to the University of Georgia, and by NSF grant
AST-9417242, NASA grant NAG5-3505 and an IBM SUR grant to the University of
Oklahoma.  Some of the calculations presented in this paper were performed on
the IBM SP2 and SGI Origin 2000 of the UGA UCNS, at the San Diego Supercomputer
Center (SDSC) and at the National Center for
Supercomputing Applications (NCSA), with support from the National Science 
Foundation, and at the NERSC with support from the DoE. We thank all these
institutions for a generous allocation of computer time.

% That's the end of the main body of the paper.  Now we will have some
% back matter.
%
% Now comes the reference list.  In this document, we used \cite to call
% out citations, so we must use \bibitem in the reference list, which
% means we use the LaTeX thebibliography environment.  Please note that
% \begin{thebibliography} is followed by a null argument.  If you forget
% this, mayhem ensues, and LaTeX will say "Perhaps a missing item?" when
% you run it.  Do not call us, do not send mail when this happens.  Put
% the silly {} after the \begin{thebibliography}.
%
% Each reference has a \bibitem command to define the citation format
% to be placed in the text (in []) and the symbolic tag used for 
% cross referencing (in {}).
%
% See sample1.tex, or the AASTeX guide, for an alternative to the \cite-
% \bibitem command.

\end{document}